\documentclass[12]{article}
\usepackage[utf8]{inputenc}
\usepackage{graphicx}
\usepackage[hidelinks]{hyperref}
\usepackage{url}
\usepackage[]{algorithm2e}
\usepackage{float}
\usepackage[section]{placeins}
\usepackage[table,xcdraw]{xcolor}
\graphicspath{{Images/}}
\usepackage[a4paper, width=180mm, top=25mm, bottom=25mm, bindingoffset=6mm]{geometry}
\usepackage{fancyhdr}
\usepackage[font=small,skip=0pt]{caption}
\usepackage{svg}
\usepackage{amsmath}
\usepackage{authblk}
\usepackage{comment}
\usepackage{enumitem}

\usepackage[normalem]{ulem}
\useunder{\uline}{\ul}{}
\usepackage{placeins}

\definecolor{darkg}{RGB}{0, 110, 0}

\date{}
\title{\huge
	{Shape-driven deep neural networks for fast acquisition of aortic 3D pressure and velocity flow fields}\\[0.5in]
}
\author{Endrit Pajaziti \thanks{endrit.pajaziti.13@ucl.ac.uk}}
\author{Javier Montalt-Tordera}
\author{Claudio Capelli}
\author{Raphael Sivera}
\author{Emilie Sauvage}
\author{Silvia Schievano}
\author{Vivek Muthurangu}

\affil[]{Institute of Cardiovascular Science, University College London}
\affil[]{*Corresponding author email: endrit.pajaziti.13@ucl.ac.uk}

\providecommand{\keywords}[1]
{
  \small	
  \textbf{\textit{Keywords---}} #1
}

\begin{document}

\maketitle

\begin{center}
\end{center}

\begin{abstract}
Computational fluid dynamics (CFD) can be used to simulate vascular haemodynamics and analyse potential treatment options. CFD has shown to be beneficial in improving patient outcomes. However, the implementation of CFD for routine clinical use is yet to be realised. Barriers for CFD include high computational resources, specialist experience needed for designing simulation set-ups, and long processing times. The aim of this study was to explore the use of machine learning (ML) to replicate conventional aortic CFD with automatic and fast regression models. Data used to train/test the model consisted of 3,000 CFD simulations performed on synthetically generated 3D aortic shapes. These subjects were generated from a statistical shape model (SSM) built on real patient-specific aortas (N=67). Inference performed on 200 test shapes resulted in average errors of 6.01\% $\pm$3.12 SD and 3.99\% $\pm$0.93 SD for pressure and velocity, respectively. Our ML-based models performed CFD in $\sim$0.075 seconds (4,000x faster than the solver). This proof-of-concept study shows that results from conventional vascular CFD can be reproduced using ML at a much faster rate, in an automatic process, and with high accuracy.
\end{abstract}

\section{Introduction}
Computational fluid dynamics (CFD) has significant potential in cardiovascular settings, with applications including modelling of complex flow patterns and non-invasive estimation of vascular pressure \cite{CFD_in_Cardiovascular_overview_Morris2016}. Importantly, several studies have validated CFD measurements against conventional clinical methods such as 4D cardiovascular magnetic resonance (CMR) imaging or catheter-based measurements \cite{CFD_validation_Biglino2015} \cite{CFD_validation_Zhu2018}. However, the true strength of CFD lies in situations where it is difficult to acquire data with conventional clinical means. For example, some studies have used CFD to assess the haemodynamic response to exercise as an alternative to cardiac catheterisation with pharmacological stress testing \cite{CFD_in_COA_Schubert2020} \cite{CFD_coarc_LaDisa_exercise}. Another example is the use of CFD for predicting the haemodynamic response to aortic stenting in patients with coarctation \cite{CFD_preop_for_COA}. Thus, CFD can minimize the need for complex invasive procedures, while improving patient-specific decision making.

Despite potential benefits, CFD is still not integrated into routine clinical practice. This is mainly due to long computation times, the requirement for large amounts of processing power, and the need for an experienced engineer to set-up simulations correctly. Unfortunately, these limitations make CFD incompatible with current clinical workflows \cite{CFD_how_to_use_models_Huberts2018}. Recently, machine learning (ML) models have been successful in replacing time consuming or computationally intensive tasks, such as medical image segmentation \cite{ML_auto_segmentation_of_RV}. In a similar fashion, the use of ML for speeding up cardiac CFD tasks is also becoming increasingly studied \cite{CFD_ML_centerline_Yevtushenko} \cite{cfd_ML_nature_reports}. The main challenges of training ML models using CFD data include: (i) poor availability of clinical data, (ii) unstructured meshes without point correspondence, and (iii) large meshes and resultant CFD flow fields. To overcome these problems, we used statistical shape modelling and dimensionality reduction techniques to produce dimensionality-reduced representations of both aortic shape and flow fields, and to enable creation of large amounts of synthetic training data \cite{ML_CFD_WeiSun} \cite{CFD_SSM_Ao_Valve}. This approach relies on the fact that in the past it has been show hemodynamic flow structures can be regressed from shape features \cite{cfd_steinman}. In this study, our objectives were: (i) to create a large synthetic cohort of 3D aortas based on real clinical cases, (ii) to train ML models to predict aortic pressure and velocity fields by representing unstructured/large data types with low-dimensional vectors, and (iii) to compare results between our fast and automatic ML-based CFD solution and our conventional CFD method.


\section{Materials \& Methods}

\subsection{Statistical Shape Modelling}
\begin{figure}[h!]
	\begin{center}
		\includegraphics[width=110mm]{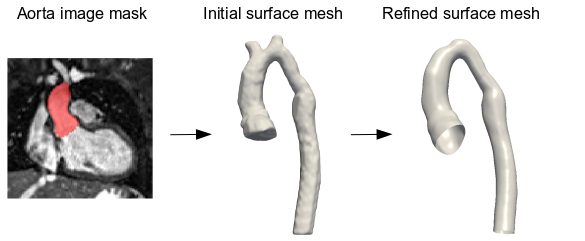}\\
	\end{center}
	\caption{The automatic pipeline for pre-processing each patient 3D model before statistical shape modelling is performed. Segmentation of the images is followed by remeshing, smoothing and clipping of inlets/outlets and head \& neck vessels.}
	\label{fig:segmentation}
\end{figure}

The dataset used for development of the statistical shape model (SSM) consisted of cardiac and respiratory gated steady state free precession CMR images (N=67) from patients previously diagnosed with coarctation of the aorta (CoA). All patients were post-surgical repair, asymptomatic and underwent CMR imaging at a mean age of 22.4 ±6.2 years. Images for each subject were segmented and converted into surface meshes (Figure \ref{fig:segmentation}). This was followed by remeshing and smoothing using functions from the vascular modelling toolkit (VMTK) \cite{VMTK}. All geometries were automatically registered using an iterative closest point algorithm in VMTK \cite{ICP_registration}. Surfaces were manually clipped above the aortic root for the inlet, and at the diaphragm for the outlet. An SSM was then built using these clipped aortic surfaces.

\begin{figure}[h!]
	\begin{center}
		\includegraphics[width=150mm]{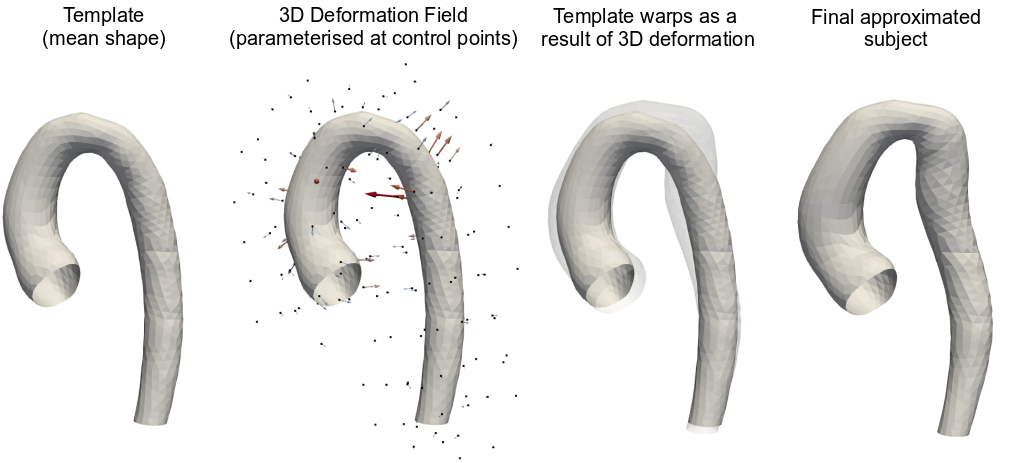}\\
	\end{center}
	\caption{An example of an approximated aortic shape using our SSM. Individual surface or volumes can be reconstructed using a mean shape and applied deformation field, as shown. The locations of all control points (n=172) is displayed.}
	\label{fig:atlas}
\end{figure}

Our estimated statistical shape model is characterised by a computed template shape (average aorta) and non-linear deformations which map the template to each individual subject \cite{ssm_jan_bruse}. Three-dimensional deformations are parameterised by infinitesimal displacements (momenta) defined at control points distributed in 3D Euclidean space (Figure \ref{fig:atlas}). Deformetrica 4 was used to build the SSM and optimise the location/number of control points \cite{deformetrica_bone2018}. The momenta from our SSM formed a 3D matrix of size [67, 172, 3], where 67 = number of subjects, 172 = number of control points, and 3 = dimensions of the displacement. In Deformetrica, non-rigid registration (deforming the template towards a target) is performed using a Gaussian kernel with a regularisation term in the loss function, in order to maximise the "smoothness" of the deformation \cite{deformetrica_bone2018}. Therefore, deformed meshes are assumed to have nodes in 'spatially correspondent' regions of the aorta (see appendix). Principal component analysis (PCA) was used to reduce the dimensionality of the momenta, meaning that each aortic deformation field (and therefore each shape) could be expressed using only 35 parameters. This equates to an almost 15 times reduction in deformation matrix size, while maintaining 99\% of the variance. The surface template computed by our SSM contained 2,541 nodes (Figure \ref{fig:atlas}). This surface template was also converted into a volume mesh with tetrahedral elements (29,000 nodes).

Using the SSM, new aortic shapes were generated by randomly sampling the previously computed PCA shape mode coefficients using a Gaussian distribution. With this approach, 3000 deformation fields were created (one for every new subject). By applying these new 3D spatial deformations onto the aortic surface and volume template, 3000 surface/volume meshes were generated. Additionally, since all deformed meshes derive from the template mesh, all synthetic subjects contained the same numbers of nodes/elements. This was vital for enabling the dimensionality reduction of derived flow fields as described in later sections. The mean centreline lengths and diameters (computed using the ”sphere-inscribed radius” function in VMTK) of the generated synthetic aorta population were compared to those of the real patient cohort.

\subsection{Computational Fluid Dynamics Pipeline}
\label{CFD_subsection}
\begin{figure}[h!]
	\begin{center}
		\includegraphics[width=140mm]{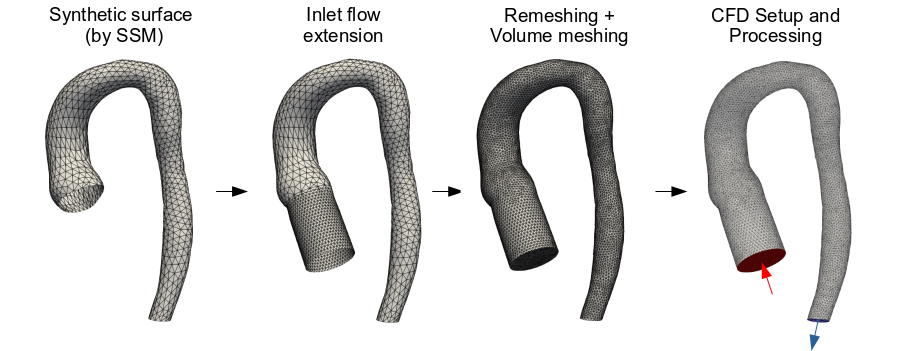}\\
	\end{center}
	\caption{Pipeline to CFD - surfaces have flow extensions added before volume meshing. CFD is performed on the final volume mesh for each synthetic subject.}
	\label{fig:cfd_pipeline}
\end{figure}

Volume meshes previously generated with the SSM were unsuitable for CFD computation. This was primarily because low mesh skewness could not be guaranteed, and remeshing was not an option since nodes were to be preserved in order to maintain point correspondence. Therefore, separate meshes solely for CFD computation were built, starting from the surface of each aorta. Firstly, each surface was extended by 40mm at the inlet (Figure \ref{fig:cfd_pipeline}). This was done to produce a flat and circular inlet upon which a velocity profile could be uniformly applied. Extending the inlet further than 40mm was avoided in order to reduce the likelihood of surface self-intersection. Following this, volume meshing with tetrahedral elements was performed \cite{tetgen} (~400,000 cells on average). Element/node counts were deemed to be in satisfactory ranges after a mesh sensitivity analysis (see appendix).

CFD (Fluent, Ansys Technologies) was performed on all 3,000 synthetic cases. The same boundary conditions were applied to each simulation, as part of adopting a simple model which would reliably converge for all subjects. Laminar, steady-state flow conditions were enforced. An inlet velocity of 1.3m/s, corresponding to an average ascending aortic flow rate at peak systole, was set \cite{cfd_aorta_velocity_in} \cite{cfd_aorta_velocity_in2}. Outlet gauge pressure was fixed at 0 Pa. Standard non-slip conditions were applied at the wall, and the fluid was assumed to be Newtonian with density and dynamic viscosity equal to 1,060 kg/m\textsuperscript{3} and 0.004 Pa$\cdot$s, respectively \cite{cfd_blood_conditions}. The set-up and simulation of all 3000 cases was fully automated.

\subsection{Machine Learning}

\subsubsection{Data Interpolation and Principal Component Analysis (PCA)}
\begin{figure}[h!]
	\begin{center}
		\includegraphics[width=110mm]{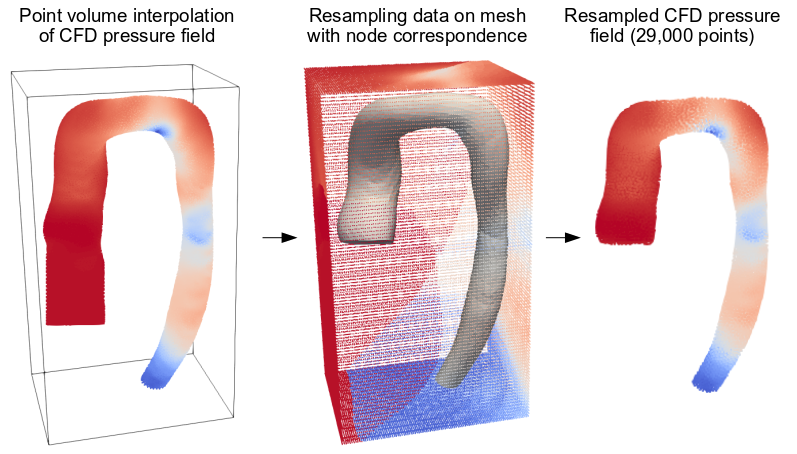}\\
	\end{center}
	\caption{Process of interpolating CFD results from CFD mesh onto a point-correspondence mesh (generated by the SSM). Node concordance is restored between subjects.}
	\label{fig:interpolation}
\end{figure}

As CFD was performed on large unstructured meshes with inconsistent numbers of nodes/elements between cases, point correspondence had to be restored prior to PCA-based dimensionality reduction of the flow fields (needed for easier model training). This was done using the volume meshes previously generated with the SSM by 'shooting' on the template. Since each of these 'SSM volume meshes' inherited its properties from the template, relative nodal positions were preserved. Consequently, pressure/velocity data for all subjects was re-sampled from unstructured CFD meshes onto SSM volume meshes using a Voronoi kernel in the software Paraview (Figure \ref{fig:interpolation}). This resulted in 3,000 newly resampled pressure/velocity fields, with all subjects containing 29,000 nodes in point correspondence. The data was concatenated into a feature vector and PCA was applied to reduce dimensionality. We aimed to capture 99\% of variance with as few PCA modes as possible.

\subsubsection{Deep Neural Network Architecture}
\begin{figure}[h!]
	\begin{center}
		\includegraphics[width=140mm]{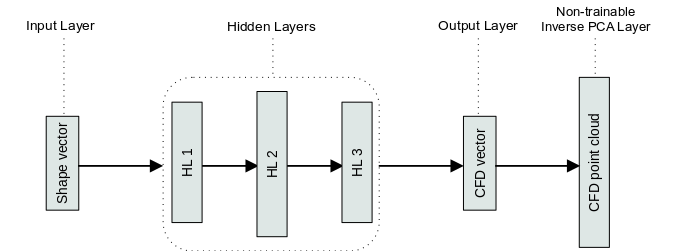}\\
	\end{center}
	\caption{The general sequential, fully-connected DNN set-up used to build both pressure and velocity predictors ('CFD vector' can be either pressure or velocity PCA vectors).}
	\label{fig:dnn_architecture}
\end{figure}

The architecture we adopted was a standard sequential, fully-connected deep neural network (DNN) with independent networks for pressure and velocity. The inputs for the model were the PCA scores of shape (reduced order deformations), which is also referred to as a ’shape vector’. The output of the trainable part of model were the pressure/velocity PCA scores (reduced order CFD field), referred to as a 'pressure/velocity vector'. A non-trainable inverse PCA layer (implemented in Keras using a lambda layer) serves to reconstruct the pressure/velocity vector into the full 3D flow field with 29,000 nodes (see Figure \ref{fig:dnn_architecture}). Rectified linear units (ReLU) were used in each hidden layer. Linear activation functions were set at the output. Model implementation was done using Keras and TensorFlow 2.0.

\subsubsection{Deep Neural Network Training}
Models were built separately for predicting the static pressure and the velocity-magnitude. The loss function used for training was the mean absolute error (MAE), computed on the entire 3D flow field (i.e. after inverse PCA) rather than on the output pressure/velocity vector. This provides a more granular measure of error and effectively weights the importance of each PCA mode in the network according to the amount of variance it explains. Model optimisation was carried out using the Adam optimiser \cite{adam_opt}. Hyperparameter tuning was conducted with a 5-fold cross-validation process to find the best model settings. The number of hidden layers, number of hidden layer neurons and initial learning rate were all explored. This was repeated for 1,000 model configurations, sampled using a tree-structured Parzen estimator (TPE) algorithm \cite{hyperparam_search}. Batch size and epochs were set at 32 and 50, respectively. Hyperband pruning was used to terminate early training rounds if the model was deemed to be poorly fitting the validation data.

After completing hyperparameter tuning using 5-fold cross-validation, the model was retrained on the entire training dataset. From 3,000 subjects, 2,800 were randomly selected to make up the training set (with 10\% being used for validation). The remaining subjects (200) formed the test set. Model training lasted 1000 epochs and training/validation loss was monitored. A workstation with an Nvidia GTX 1080Ti graphics card was used to perform training.

\subsubsection{Model Evaluation}

Once trained, models were evaluated on the test set of synthetic aortas (n=200). Absolute errors were computed for every node in all test cases by comparing the prediction value (ML) to the ground-truth value (CFD). Errors was then normalised according to subject CFD data range, as detailed in Liang et. al \cite{ML_CFD_WeiSun}. Normalisation was necessary to enable direct comparison between individual cases and also between CFD metrics, since pressure/velocity ranges widely differed per subject. Formula \ref{NAE} details how normalised absolute error (NAE) is computed for pressure or velocity at a node \(i\), belonging to a subject \(j\). \(True\) and \(Pred\) refer to CFD and ML values, respectively.

\begin{equation}
    NAE(i,\;j) = \frac{\left| True_{i,\;j}\; - \; Pred_{i,\;j} \right|}{Range(True_{j})} \times100\%
    \label{NAE}
\end{equation}\\

Mean node errors (MNAE\textsubscript{N}) were computed by averaging NAE values across the population for each node (n=29,000). These were then plotted on the template mesh points in order to better visualise the magnitude of these errors with respect to their location. However, since NAE values are absolute errors this provides no insight regarding model bias. Therefore, a Bland-Altman plot was used to examine the bias and limits of agreement of the pressure and velocity DNNs. This was done for the overall aorta and for three separate regions; ascending aorta, transverse arch and descending aorta (anatomically defined).

Mean subject errors (MNAE\textsubscript{S}) were computed by averaging NAE values in each subject (n=200). Cases with the best, median and worse mean subject error values were compared. A single population error for both pressure and velocity was given by averaging all MNAE\textsubscript{S} values. The relationship between shape mode scores and subject error (MNAE\textsubscript{S}) was investigated with scatter plots and assessed using Pearson R coefficients and p values (p=0.05 considered significant).

\begin{figure}[h!]
	\begin{center}
		\includegraphics[width=70mm]{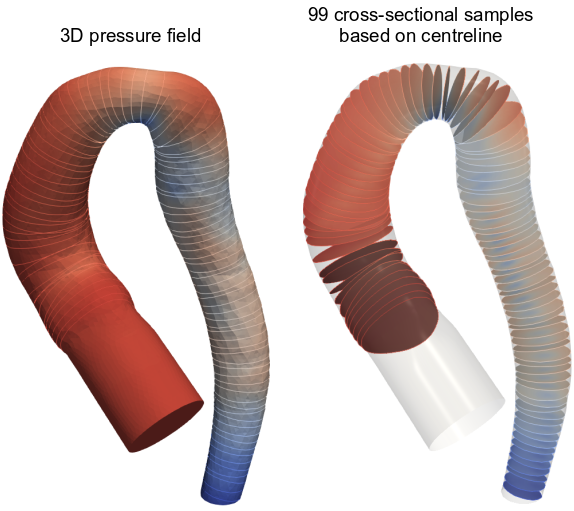}\\
	\end{center}
	\caption{Method used to sample cross-sectional pressure/velocity gradients from 3D flow fields. Pressure/velocity is averaged over each plane. The origin is always the aortic root (excluding the extension). Final gradient pressure/velocity arrays have 99 points.}
	\label{fig:plane_sampling}
\end{figure}

In addition to node-based comparison between CFD and ML flow fields, an approach to compare more conventional pressure/velocity gradients was also used. In order to compute gradients, subject centrelines were used to create cross-sectional planes at 99 locations along the length of each aortic 3D flow field. Average pressure and velocity values were then sampled at each plane (Figure \ref{fig:plane_sampling}). To compare gradients, a common regression error metric was used (mean absolute percentage error) \cite{mape_error_metric}. This metric is unsuitable when true values may lie very close to 0 (which is the case in the 3D velocity fields). Hence, mean absolute percentage error (MAPE) was only used for computing gradient errors.

\begin{equation}
    MAPE = \frac{1}{N}\sum_{i=1}^{N}\left| \frac{True_{i}\;-\;Pred_{i}}{True_{i}} \right| \times100\%
    \label{MAPE}
\end{equation}

Finally, the models were tested on 10 real, patient-specific aortas with previously repaired CoA, completely unseen from the SSM and the DNN. This was done in order to validate the robustness of the models for inferring accurate flow fields on real subjects outside of our synthetic training/testing sets. CMR images of each patient were segmented. Surfaces were approximated by non-rigid registration (applying deformations on the template) using the SSM. Deformation matrices for each case were decomposed into PCA shape vectors and passed as inputs into the DNN models. Pressure and velocity-magnitude fields were inferred for each subject. Following this, all ten predictions were compared to their corresponding, ground-truth CFD flow fields using the MNAE\textsubscript{S} and MAPE metrics, as previously described.


\section{Results}

\subsection{Statistical Shape Modelling}
\begin{figure}[h!]
	\begin{center}
		\includegraphics[width=170mm]{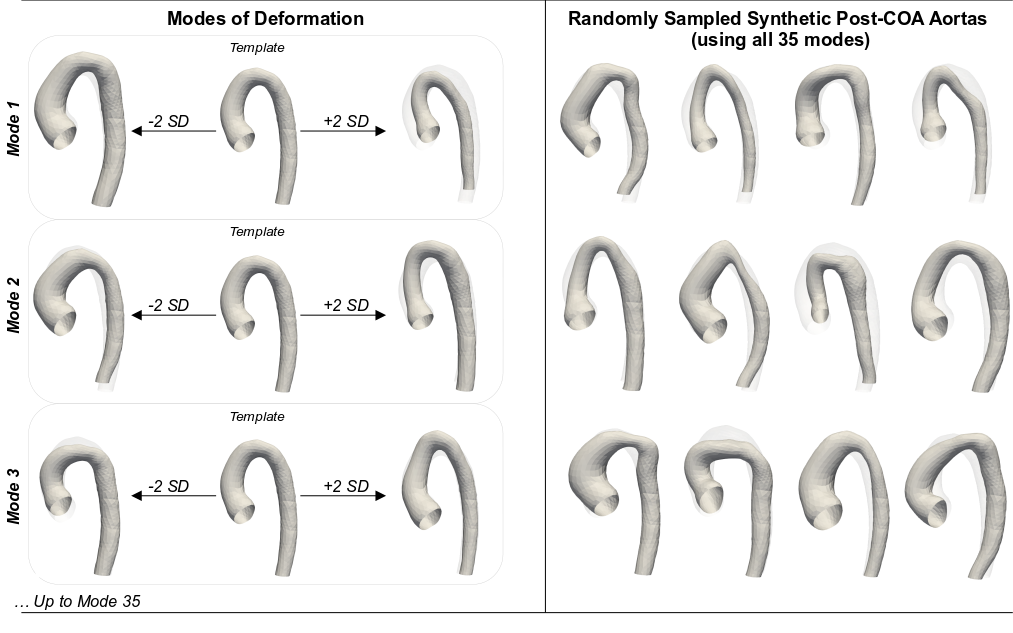}\\
	\end{center}
	\caption{Left: first three modes of deformation from the SSM (SD = standard deviation). Right: examples of synthetic post-CoA aortas from the test set (using combinations of all 35 shape modes).}
	\label{fig:ssm_shapes}
\end{figure}

PCA decomposition was performed on the deformation matrices (momenta) computed by statistical shape modelling. The first, second and third PCA modes captured 29.8\%, 13.2\% and 10.1\% (total 53.1\%) of the variability, respectively. Mode 1 relates to overall vessel size, mode 2 relates to ascending arch angulation/diameter and mode 3 describes how 'gothic' the aorta is (characterised by a more triangular arch). After PCA decomposition, 99\% of the variance in the momenta could be represented with the first 35 modes. Some examples of the 3,000 synthetic subjects produced by randomly sampling and combining 35 PCA mode scores are shown in Figure \ref{fig:ssm_shapes}. Characteristic dimensions of the synthetic aortas (length and diameter) were found to be statistically similar to those of the real patient cohort (see Table \ref{tab:table_centerline}).

\begin{table}[h!]
\resizebox{\columnwidth}{!}{%
\begin{tabular}{lrrrr}
\hline
 & Mean aortic vessel length (mm) & Mean aortic vessel diameter (mm) \\ \hline
Real patient cohort (n=67) & 257.3 $\pm$ 29.88 SD & 19.74 $\pm$ 1.29 SD \\
Synthetic patient cohort (n=3000) & 258.5 $\pm$ 23.08 SD & 20.12 $\pm$ 1.32 SD \\
p-value (Welch's t-test) & 0.75 & 0.25 \\ \hline
\end{tabular}
}
\caption{\label{tab:table_centerline} Comparison of aorta dimensions between original real cohort (n=67) and synthetic cohort (n=3000).}
\end{table}

\subsection{Training Data}
After CFD was computed on all cases (n=3000), values were interpolated from high resolution meshes onto lower-resolution grids in point correspondence. Interpolation errors (MAPE) were computed by comparing pressure/velocity gradients in each case (Figure \ref{fig:plane_sampling}). The mean loss in accuracy due to interpolation was found to be 0.056\% $\pm$0.027 and 0.849\% $\pm$0.247 for pressure and velocity-magnitude, respectively. The data post-interpolation was used as the 'ground-truth' training and testing sets. 

\begin{figure}[H]
	\begin{center}
		\includegraphics[width=150mm]{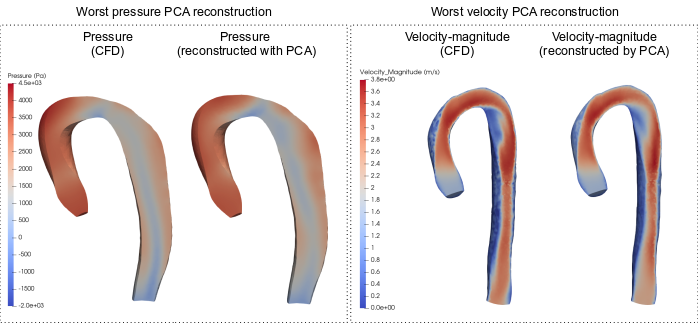}\\
	\end{center}
	\caption{Distribution of errors due to PCA. Left: subject with highest errors in pressure after reconstruction with 20 PCA modes (4.32\%). Right: subject with highest errors in velocity after reconstruction with 55 PCA modes (4.93\%). Negative pressure values are respective to the reference gauge pressure set during simulations (atmospheric pressure, 100kPa).}
	\label{fig:PCA_loss}
\end{figure}

PCA decomposition of the pressure and velocity training data matrices (2,800 x 29,000 each) was then performed, following standardisation. After PCA decomposition, 99\% of the standardised pressure variance could be captured with 20 modes. Only 87\% of the standardised velocity variance could be captured with 55 modes, and it was felt that adding more modes to capture greater variance was not feasible due to massively diminishing returns. Subject errors resulting from PCA decomposition were tested on the 200 test cases (unseen by the PCA model). Average MNAE\textsubscript{S} in pressure and velocity fields were found to be 1.46\% $\pm$ 0.59 SD and 2.70\% $\pm$0.49 SD. The reconstructed test cases with the highest MNAE\textsubscript{S} for pressure and velocity (4.32\% and 4.93\%, respectively) are shown in Figure \ref{fig:PCA_loss}. 

\subsection{Model Architecture}
\begin{figure}[H]
	\begin{center}
		\includegraphics[width=170mm]{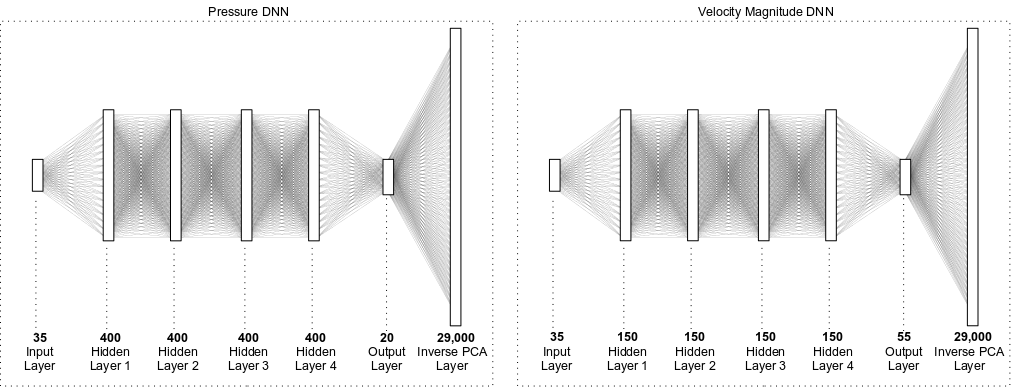}\\
	\end{center}
	\caption{The architecture and parameters of both pressure and velocity-magnitude networks after training.}
	\label{fig:dnn_params}
\end{figure}

The model input layer size was set at 35 (number of shape modes). Output layer sizes were set at 20 and 55 for pressure and velocity, respectively (number of pressure/velocity modes). Hyperparameter tuning using cross-validation was performed 1,000 times to search for the optimal learning rate, number of neurons and number of layers, with tuning taking $\sim$4 hours per model. Pressure and velocity model architectures as a result of the optimisation process are shown in Figure \ref{fig:dnn_params}.

\subsection{Model Predictive Performance}
Pressure and velocity-magnitude fields were computed on the test set (n=200) using the trained DNNs. Inference took ~0.075 seconds per subject for both DNNs. In comparison, conventional CFD took $\sim$5 minutes on average for convergence, demonstrating an approximate 4,000x speed-up with ML.

\subsubsection{Node errors}

\begin{figure}[h!]
	\begin{center}
		\includegraphics[width=100mm]{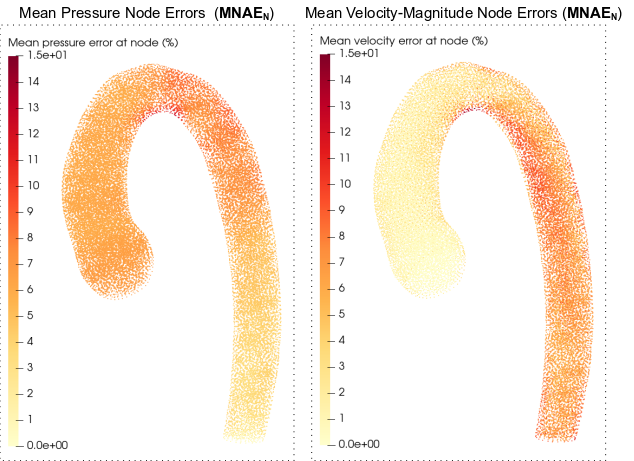}\\
	\end{center}
	\caption{Distribution of mean errors for all nodes (MNAE\textsubscript{N}), computed on test set (n=200). Values are projected on the template volume mesh (average position of the nodes) for visualisation.}
	\label{fig:node_error}
\end{figure}

Average node prediction errors (MNAE\textsubscript{N}) were projected onto the template to visually assess the distribution of errors (Figure \ref{fig:node_error}). Pressure errors were observed to be larger in the ascending aortic region. Velocity errors were more prevalent in the underside of the arch and descending aorta. Bland-Altman analysis showed negligible prediction biases for the overall aorta (Figure \ref{fig:bland-altman} and within selected regions (see appendix). 

\begin{figure}[h!]
	\begin{center}
		\includegraphics[width=180mm]{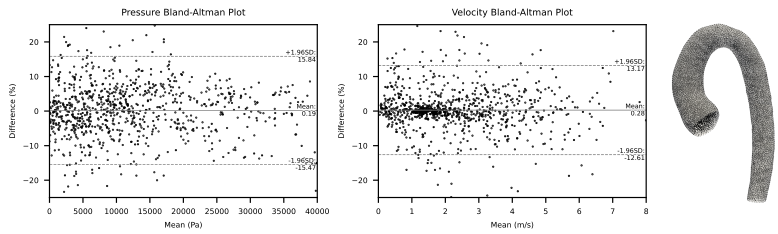}\\
	\end{center}
	\caption{Bland-Altman plot for evaluating bias and limits of agreement for DNN models. All nodes (n=29,000) for each subject (n=200) were concatenated (5,800,000 data points). Each value was compared between ML and CFD values. Only 1,000 randomly selected points were graphically drawn to improve readability.}
	\label{fig:bland-altman}
\end{figure}

\subsubsection{Subject errors}

\begin{figure}[h!]
	\begin{center}
		\includegraphics[width=180mm]{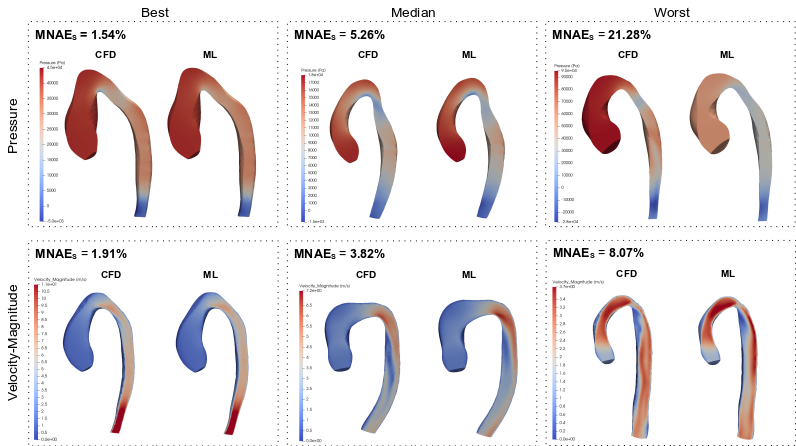}\\
	\end{center}
	\caption{Comparisons between ground truth (CFD) and predicted (ML) in the test set (n=200). Best, median and worst cases for both pressure and velocity-magnitude are shown.}
	\label{fig:dnn_prediction}
\end{figure}

\begin{figure}[h!]
	\begin{center}
		\includegraphics[width=170mm]{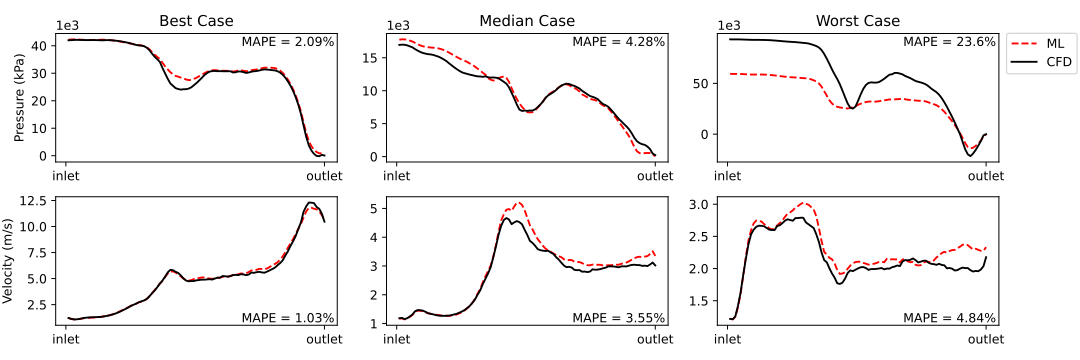}
	\end{center}
	\caption{Comparison of pressure/velocity gradients for the best, median and worst cases in the test set (n=200) are shown. Subjects are the same as those in Figure \ref{fig:dnn_prediction}. Gradient errors (MAPE) between ML and CFD for each subject are displayed.}
	\label{fig:gradient errors}
\end{figure}

\begin{figure}[h!]
	\begin{center}
		\includegraphics[width=130mm]{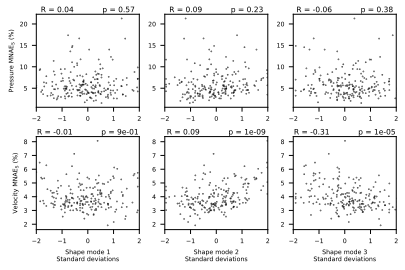}\\
	\end{center}
	\caption{Scatter plot comparing the subject error (MNAE\textsubscript{S}) against shape PCA mode values in the test set (n=200). Pearson~R coefficients and p-values were computed for each subplot.}
	\label{fig:shape_vs_error}
\end{figure}

The population error for pressure and velocity was 6.01\% $\pm$3.12 SD and 3.99\% $\pm$0.93 SD, respectively. The test cases with the best, median and worst subject error (MNAE\textsubscript{S}) are shown in Figure \ref{fig:dnn_prediction}. The gradients of pressure/velocity for these same cases was also computed (Figure~\ref{fig:gradient errors}). The relationship between pressure/velocity MNAE\textsubscript{S} and shape mode coefficients is presented in Figure \ref{fig:shape_vs_error}. The second and third mode showed statistically significant correlations with velocity prediction MNAE\textsubscript{S}, with the third shape mode showing the highest correlation (R=-0.31). No significant correlations were found between any other shape mode and MNAE\textsubscript{S}.

\subsection{Validation with real patient data}
\begin{figure}[h!]
	\begin{center}
		\includegraphics[width=180mm]{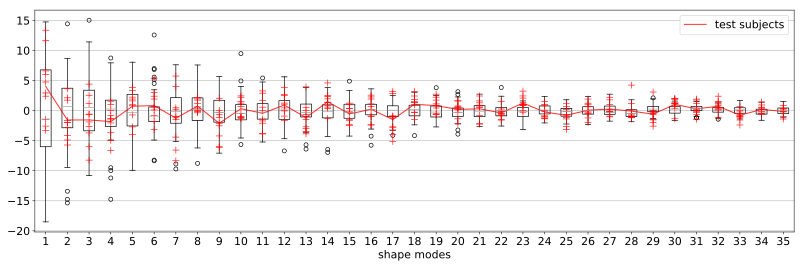}\\
	\end{center}
	\caption{Projection of new prospective test cases (n=10) on all 35 SSM shape modes. Boxplots represent the range of PCA mode scores in the original cohort (n=67). Whiskers are 1.5 $\times$ the interquartile range. Red markers are the PCA scores for the new subjects (n=10).}
	\label{fig:pca_projection}
\end{figure}

\begin{figure}[h!]
	\begin{center}
		\includegraphics[width=170mm]{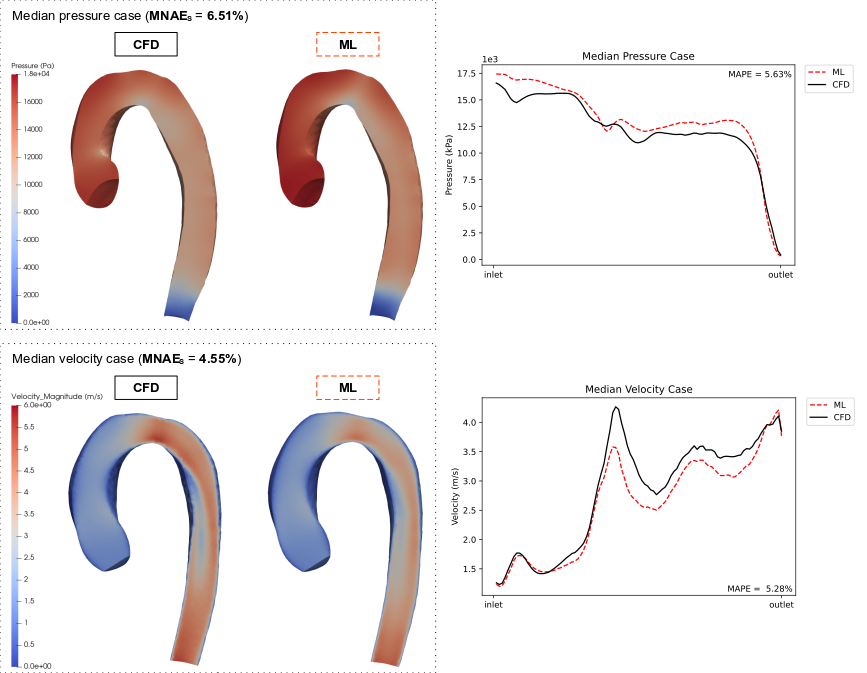}\\
	\end{center}
	\caption{Median pressure and velocity predictions on the real patient test cohort (n=10). CFD (ground truth) and ML flow fields are shown. Pressure/velocity gradients for these cases are also displayed, including a gradient error (MAPE).}
	\label{fig:DNN_pred_prospective}
\end{figure}

Ten patient cases with previously repaired coarctation of the aorta were collected and segmented. Surfaces were approximated through non-rigid registration (deformation of the SSM template mesh). The corresponding deformations were decomposed into shape PCA vectors. Shape PCA projections were found to be in acceptable ranges when compared to those from the original cohort (Figure \ref{fig:pca_projection}). Shape PCA vectors were then inputted into both pressure and velocity DNNs. Predictions were then compared against CFD simulations performed using the same pipeline as previously described (Page \ref{CFD_subsection}). The population error was found to be 10.19\% $\pm$10.41 and 4.47\% $\pm$1.18 for pressure and velocity, respectively (n=10). This corresponds to an increase in mean subject error by 4.18\% for pressure and 0.48\% for velocity, when compared to the test set of 200 synthetic aortas. The subjects with the median pressure and velocity MNAE\textsubscript{S} are shown in Figure~\ref{fig:DNN_pred_prospective}.

\section{Discussion}
The main findings of this study were: i) statistical shape models (SSMs) and PCA are suitable for creating synthetic training data and dimensionality-reduced representations of 3D shape and flow, ii) DNNs based on these dimensionality-reduced representations can predict pressure and velocity fields with high accuracy.

\subsection{Synthetic Data Generation and Dimensionality Reduction}
A key element of our approach was to use an SSM and PCA to both generate a synthetic training dataset (n=3,000), and to parameterise aortic shape/flow fields for simplifying DNN model training. We demonstrated that less than 60 PCA modes are needed to capture the majority of variance within aortic shape deformations and 3D pressure/velocity fields. Mean PCA reconstruction errors of pressure/velocity were found to be low, but not insignificant in the worst observed test-set reconstructions (Figure \ref{fig:PCA_loss}). Geometric properties of our synthetic data were shown to be mostly close to the real cohort (Table \ref{tab:table_centerline}). In the future, newer approaches for generating synthetic data and creating dimensionality-reduced representations of complex structures should be explored, notably deep-learning methods such as autoencoders \cite{ML_CFD_WeiSun} and generative adversarial networks (GANs) \cite{GANs}. In some studies, autoencoders have been shown to be superior to PCA-based methods (e.g. for 3D facial surface reconstruction) \cite{ssm_pca_vs_autoencoder} \cite{ML_autoencoders}. Additionally, GANs have shown promise for creating images of synthetic patients afflicted with congenital heart disease \cite{ML_generation_of_synthetic}. Finally, instead of sampling a Gaussian distribution to find combinations of shape PCA parameters, other methods for creating new DNN training data may be more appropriate such as Latin hypercube sampling \cite{latin_hypercube_sampling}.

\subsection{Deep Neural Networks}
Our ML models were observed to predict point clouds of both pressure and velocity flow fields with good accuracy, while being approximately 4,000x faster than our conventional CFD method. Node errors for pressure were seen to be larger at the inlet, while velocity errors were more skewed towards the distal regions of the aorta (Figure \ref{fig:node_error}). This is most likely due to the CFD boundary conditions at the inlet and outlet constraining the pressure/velocity variability at these regions. The further away from the aortic inlet or outlet, the greater the variability in velocity or pressure, respectively. It should be noted that there were no significant biases in either pressure or velocity predictions, suggesting that there were no systematic errors with the models (Figure \ref{fig:bland-altman}). In testing, we found that the mean pressure subject error was slightly higher than the mean velocity error. This is despite the pressure PCA model capturing more variance than the velocity model. A possible explanation for the higher pressure errors is that the association between shape and pressure is more complex than that between shape and velocity in aortic domains. This is supported by the observation that shape modes do not correlate with pressure errors (Figure \ref{fig:shape_vs_error}). Interestingly, there was a strong negative correlation between shape mode 3 and velocity error. This suggests that more ’gothic’ aortas (characterised by a more triangular arch) were less prone to velocity prediction errors. A possible explanation may be that the gothic arch constrains downstream flow patterns (where most velocity errors occur), hence making it easier for the model to characterise flow features associated to this subset of aortic shapes. 

The approach of using DNNs to model 3D aortic pressure and velocity flow fields has been described in other works \cite{ML_CFD_WeiSun} \cite{cfd_ML_nature_reports}. In this study, fully-connected DNNs were used. Other studies have used recurrent neural networks or specialised architectures such as PointNet to build CFD-based ML models \cite{CFD_ML_centerline_Yevtushenko} \cite{CFD_ML_ships_pointnet} \cite{pointnet}. Further DNN architectures should be tested in the future, including probabilistic architectures which enable outputting uncertainty intervals during inference \cite{CFD_ML_autodesk_cars} \cite{CFD_ML_ships_pointnet} \cite{ML_Bayes_backprop}. In order to improve model performance, more training data should be generated, which will in turn enable the use of more complex models.


\section{Limitations}
In order to translate our DNN-based CFD approach to clinics, there are two main modelling limitations which need to be bypassed. The first is related to the loss in surface accuracy when using SSM representations of aortic shapes. The second is the current simplicity of the CFD approach, which needs to be further developed in order to generate more meaningful DNN training data.

\subsection{Shape Parameterisation}
It has been seen in previous studies that aortic CFD flow fields are highly sensitive to geometric and topological variation \cite{cfd_geom_autoseg_javier_jcmr} \cite{cfd_geom_is_important_KwongTse}. For this reason, using shape vectors that are accurate descriptors of the aortic surfaces is critical. In this study, we performed ML predictions and CFD either on synthetic aortas or on SSM approximations of real aortas, both of which are completely parameterised by our shape PCA vectors. However, SSM approximations of real aortas do not match the 'true' segmentation surface one-to-one. In the future, the effect of this registration error on the resultant CFD flow fields should be investigated. Where possible, augmentations to the shape vector should be trialled in order to see if DNN prediction errors for real subjects can be reduced. This may require including additional shape information as inputs in the DNN models, such as a registration error or a geometric feature (e.g. centreline diameters). Such complimentary shape descriptors may be used to better inform the network of important features not fully captured by the shape vector alone. Additionally, multiple SSMs based upon templates other than the mean aortic shape should be generated. These would enable closer non-rigid registration for unique cases where the target aorta deviates significantly from the mean shape. Finally, more subjects should be added to our SSM in order to be able to fit a wider variety of aortic shapes.

\subsection{Computational Fluid Dynamics}
In this study, a simplified CFD pipeline was chosen in order to easily automate and ensure convergence for a large number of simulations (n=3,000). A standard CFD solver set-up was used, assuming steady-state conditions and incompressible flow. Previous studies have found this solver set-up to be sufficiently accurate for producing aortic pressure gradients in clinical applications \cite{cfd_simple_models_useful}. Additionally, it was assumed that the flow through the great arteries during systole was laminar during peak systole \cite{cfd_review_on_aorta_cfd}. In the future, Reynolds numbers may be computed for individual cases in order to properly assign turbulence models. However, for complex morphologies (such as CoA), Reynolds number has been seen to be an inconsistent measure of turbulence \cite{cfd_turbulence_coa}. Some studies propose an alternative solution to turbulence modelling when performing CFD on a large scale, which involves meshing the domain with extremely high numbers of nodes \cite{cfd_ML_nature_reports}.

Boundary conditions we selected included a fixed, flat velocity inlet and zero pressure outlet condition for all cases. However, idealised inlet velocity profiles (flat, parabolic etc.) have been shown to be ineffective for producing clinically relevant data \cite{cfd_inlet_idealised_isBad} \cite{cfd_inlet_BC_Morbiducci_idealisedIsBad}. In the future, image-derived flow profiles should be included as a DNN input parameter and as part of the synthetic CFD training data generation \cite{ssm_yevtushenko}. In this study, it was decided to omit the head and neck vessels from the CFD model for simplicity, however would be required when aiming to simulate realistic patient-specific aortic hemodynamics \cite{cfd_geom_is_important_KwongTse}. The inclusion of head and neck vessels along with lumped parameter outlet models such as Windkessel models should be explored in the future \cite{cfd_windkessel_overview} \cite{cfd_windkessel_study}, as modelling downstream resistance has been seen to produce more clinically meaningful results \cite{cfd_outlet_resistance_model_abdominal_aneurysm} \cite{cfd_inletoutlet_dont_affect_much} \cite{cfd_aorta_BC_study}.


\section{Conclusion}
In this proof-of-concept study, we have proposed a pipeline for building ML-based models to perform repetitive vessel-based CFD tasks. Generation of synthetic aortic training data by means of shape modelling allowed ML techniques to be used even where data scarcity is an issue (n=67). Point correspondence was maintained between subject meshes in order to enable PCA. Our ML models were able to compute pressure and velocity flow fields much more rapidly (4,000x) than traditional CFD solvers, without large computational requirements or simulation setup. Comparison between predicted and ground truth test cases revealed good overall performance. The approach described in this study is shape-driven and is applicable to any vascular structure which can be segmented from medical images. In the future, the models should be improved so they can perform inference on prospective data from real patients. The requirements for this are two-fold; improving the accuracy of the shape representation methods, and using a more realistic CFD pipeline for generating training data. Finally, new DNN models should be tailored for specific clinical applications (such as for predicting exercise conditions).

\section*{List of Abbreviations}
\begin{itemize}[noitemsep]
  \item DNN = deep neural network
  \item ML = machine learning
  \item CFD = computational fluid dynamics
  \item SSM = statistical shape model
  \item PCA = principal component analysis
  \item CoA = coarctation of the aorta
  \item NAE = normalised absolute error
  \item MNAE = mean normalised absolute error
  \item MAPE = mean absolute percentage error
  \item CMR = cardiac magnetic resonance
  \item CT = computed tomography
  \item VMTK = vascular modelling toolkit
  \item LoA = Limits of Agreement
  
\end{itemize}

\subsection*{Data accessibility statement}
All code used was written by EP and RS. Scripts for preprocessing data and training DNNs are publicly available to download from Github: \url{https://github.com/EndritPJ/CFD_Machine_Learning}

\subsection*{Author Contributions}
EP: meshing, ML and SSM code, data analysis, writing manuscript original draft.\\
JMT: ML methodology and manuscript revision.\\
CC: conceptualisation, supervision, manuscript revision.\\
RS: SSM methodology, SSM code, manuscript revision.\\
ES: CFD methodology, manuscript revision.\\ 
SS: conceptualisation, supervision, extensive manuscript revision.\\
VM: conceptualisation, supervision, ML methodology, extensive manuscript revision.\\
All authors have given final approval for publication. 

\subsection*{Funding}
This work was supported by UK Research and Innovation (MR/S032290/1), Heart Research UK (RG2661/17/20), the British Heart Foundation (NH/18/1/33511, PG/17/6/32797), the Engineering and Physical Sciences Research Council (EP/N02124X/1), Action Medical Research (GN2572) and the European Research Council (ERC-2017-StG-757923).

\subsection*{Conflict of Interest}
None.

\appendix

\subsection{Appendices}

\begin{figure}[H]
	\begin{center}
		\includegraphics[width=160mm]{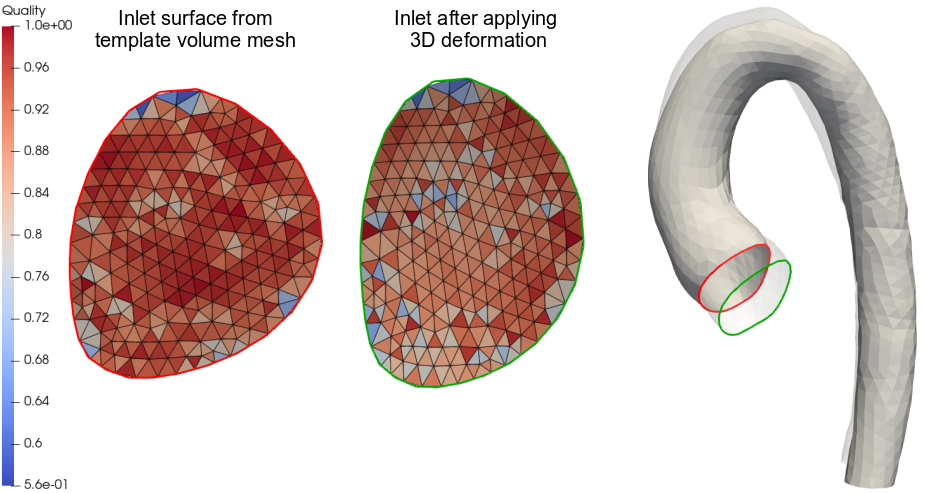}\\
	\end{center}
	\caption{Example of template volume mesh deformation for generating a new subject. The inlet surface of the template volume mesh pre-deformation (red), and the inlet post-deformation (green) are shown. Mesh skewness (Jacobian) is plotted, with a result of 1 implying optimal triangle quality. It can be observed how warping affects mesh skewness while preserving a smooth nodal distribution.}
	\label{fig:ssm_volume_deformation}
\end{figure}

\begin{figure}[H]
	\begin{center}
		\includegraphics[width=170mm]{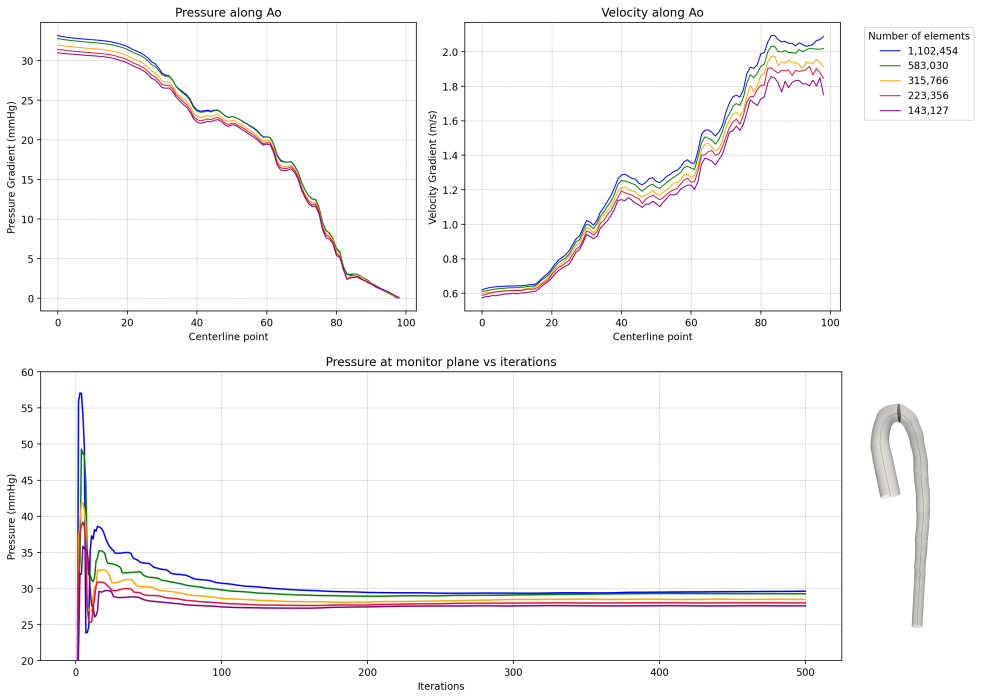}\\
	\end{center}
	\caption{CFD sensitivity analysis performed on test aortic shape with very sharp angulation. Pressure is monitored at a specific plane as well as along the entire centerline. Results show that results are comparable even with relatively low numbers of cells. Few iterations are required for convergence with the solver setup used.}
	\label{fig:sensitivity}
\end{figure}

\begin{figure}[H]
	\begin{center}
		\includegraphics[width=170mm]{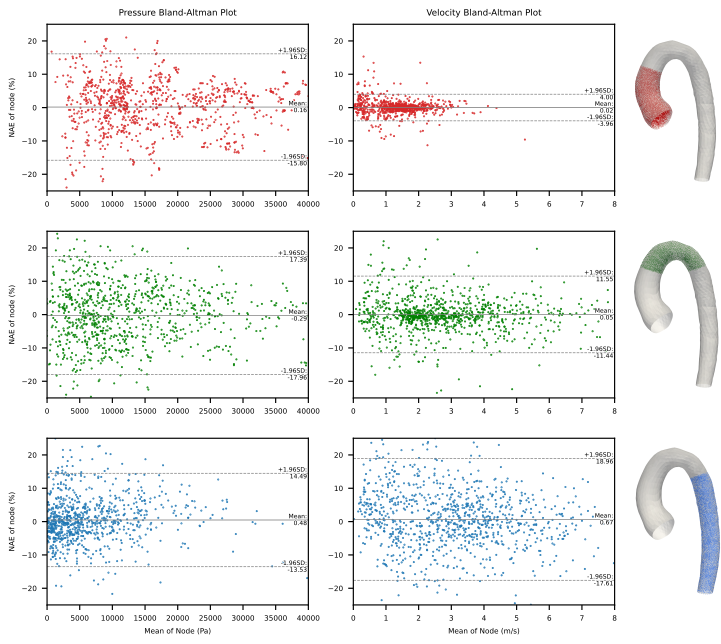}\\
	\end{center}
	\caption{Further Bland-Altman analysis comparing the different sections of the aorta. The ascending, transverse arch and descending aortic regions are all shown.}
	\label{fig:bland_altman_appendix}
\end{figure}

\bibliographystyle{vancouver}
\bibliography{main.bib}

\begin{thebibliography}{10}

\bibitem{CFD_in_Cardiovascular_overview_Morris2016}
Morris PD, Narracott A, {Von Tengg-Kobligk} H, Soto DAS, Hsiao S, Lungu A,
  et~al.
\newblock {Computational fluid dynamics modelling in cardiovascular medicine}.
\newblock Heart. 2016;102(1):18-28.

\bibitem{CFD_validation_Biglino2015}
Biglino G, Cosentino D, Steeden JA, {De Nova} L, Castelli M, Ntsinjana H,
  et~al.
\newblock {Using 4D Cardiovascular Magnetic Resonance Imaging to Validate
  Computational Fluid Dynamics: A Case Study}.
\newblock Frontiers in Pediatrics. 2015;3(December):1-10.

\bibitem{CFD_validation_Zhu2018}
Zhu Y, Chen R, Juan YH, Li H, Wang J, Yu Z, et~al.
\newblock {Clinical validation and assessment of aortic hemodynamics using
  computational fluid dynamics simulations from computed tomography
  angiography}.
\newblock BioMedical Engineering Online. 2018;17(1):1-12.

\bibitem{CFD_in_COA_Schubert2020}
Schubert C, Br{\"{u}}ning J, Goubergrits L, Hennemuth A, Berger F, K{\"{u}}hne
  T, et~al.
\newblock {Assessment of hemodynamic responses to exercise in aortic
  coarctation using MRI-ergometry in combination with computational fluid
  dynamics}.
\newblock Scientific Reports. 2020;10(1):1-12.
\newblock Available from: \url{https://doi.org/10.1038/s41598-020-75689-z}.

\bibitem{CFD_coarc_LaDisa_exercise}
LaDisa JF, Alberto~Figueroa C, Vignon-Clementel IE, Jin~Kim H, Xiao N, Ellwein
  LM, et~al.
\newblock Computational simulations for aortic coarctation: representative
  results from a sampling of patients.
\newblock Journal of Biomedical Engineering. 2011.

\bibitem{CFD_preop_for_COA}
Itu L, Sharma P, Ralovich K, Mihalef V, Ionasec R, Everett A, et~al.
\newblock Non-invasive hemodynamic assessment of aortic coarctation: validation
  with in vivo measurements.
\newblock Annals of biomedical engineering. 2013;41(4):669-81.

\bibitem{CFD_how_to_use_models_Huberts2018}
{What is needed to make cardiovascular models suitable for clinical decision
  support? A viewpoint paper}.
\newblock Journal of Computational Science. 2018;24:68-84.
\newblock Available from: \url{https://doi.org/10.1016/j.jocs.2017.07.006}.

\bibitem{ML_auto_segmentation_of_RV}
Avendi MR, Kheradvar A, Jafarkhani H.
\newblock A combined deep-learning and deformable-model approach to fully
  automatic segmentation of the left ventricle in cardiac MRI.
\newblock Medical image analysis. 2016;30:108-19.

\bibitem{CFD_ML_centerline_Yevtushenko}
Yevtushenko P, Goubergrits L, Gundelwein L, Setio A, Heimann T, Ramm H, et~al.
\newblock Deep Learning Based Centerline-Aggregated Aortic Hemodynamics: An
  Efficient Alternative to Numerical Modelling of Hemodynamics.
\newblock IEEE Journal of Biomedical and Health Informatics. 2021.

\bibitem{cfd_ML_nature_reports}
{Accelerating massively parallel hemodynamic models of coarctation of the aorta
  using neural networks}.
\newblock Scientific Reports. 2020;10(1):1-13.
\newblock Available from: \url{http://dx.doi.org/10.1038/s41598-020-66225-0}.

\bibitem{ML_CFD_WeiSun}
Liang L, Mao W, Sun W.
\newblock {A feasibility study of deep learning for predicting hemodynamics of
  human thoracic aorta}.
\newblock Journal of Biomechanics. 2020;99.
\newblock Available from: \url{https://doi.org/10.1016/j.jbiomech.2019.109544}.

\bibitem{CFD_SSM_Ao_Valve}
Hoeijmakers MJMM, Waechter-Stehle I, Weese J, {Van de Vosse} FN.
\newblock {Combining statistical shape modeling, CFD, and meta-modeling to
  approximate the patient-specific pressure-drop across the aortic valve in
  real-time}.
\newblock International Journal for Numerical Methods in Biomedical
  Engineering. 2020;36(10):1-18.

\bibitem{cfd_steinman}
Lee SW, Antiga L, Spence JD, Steinman DA.
\newblock Geometry of the carotid bifurcation predicts its exposure to
  disturbed flow.
\newblock Stroke. 2008;39(8):2341-7.

\bibitem{VMTK}
Antiga L, Piccinelli M, Botti L, Ene-Iordache B, Remuzzi A, Steinman DA.
\newblock An image-based modeling framework for patient-specific computational
  hemodynamics.
\newblock Medical \& biological engineering \& computing. 2008;46(11):1097-112.

\bibitem{ICP_registration}
Besl PJ, McKay ND.
\newblock Method for registration of 3-D shapes.
\newblock In: Sensor fusion IV: control paradigms and data structures. vol.
  1611. Spie; 1992. p. 586-606.

\bibitem{ssm_jan_bruse}
Bruse JL, McLeod K, Biglino G, Ntsinjana HN, Capelli C, Hsia TY, et~al.
\newblock A statistical shape modelling framework to extract 3D shape
  biomarkers from medical imaging data: assessing arch morphology of repaired
  coarctation of the aorta.
\newblock BMC medical imaging. 2016;16(1):1-19.

\bibitem{deformetrica_bone2018}
B{\^o}ne A, Louis M, Martin B, Durrleman S.
\newblock Deformetrica 4: an open-source software for statistical shape
  analysis.
\newblock In: International Workshop on Shape in Medical Imaging. Springer;
  2018. p. 3-13.

\bibitem{tetgen}
Hang S.
\newblock TetGen, a Delaunay-based quality tetrahedral mesh generator.
\newblock ACM Trans Math Softw. 2015;41(2):11.

\bibitem{cfd_aorta_velocity_in}
Garcia J, van~der Palen RL, Bollache E, Jarvis K, Rose MJ, Barker AJ, et~al.
\newblock Distribution of blood flow velocity in the normal aorta: effect of
  age and gender.
\newblock Journal of Magnetic Resonance Imaging. 2018;47(2):487-98.

\bibitem{cfd_aorta_velocity_in2}
Powell A, Maier S, Chung T, Geva T.
\newblock Phase-velocity cine magnetic resonance imaging measurement of
  pulsatile blood flow in children and young adults: in vitro and in vivo
  validation.
\newblock Pediatric cardiology. 2000;21(2):104-10.

\bibitem{cfd_blood_conditions}
Bonfanti M, Balabani S, Greenwood JP, Puppala S, Homer-Vanniasinkam S,
  D{\'\i}az-Zuccarini V.
\newblock Computational tools for clinical support: a multi-scale compliant
  model for haemodynamic simulations in an aortic dissection based on
  multi-modal imaging data.
\newblock Journal of The Royal Society Interface. 2017;14(136):20170632.

\bibitem{adam_opt}
Kingma DP, Ba J.
\newblock Adam: A method for stochastic optimization.
\newblock arXiv preprint arXiv:14126980. 2014.

\bibitem{hyperparam_search}
Bergstra J, Bardenet R, Bengio Y, K{\'e}gl B.
\newblock Algorithms for hyper-parameter optimization.
\newblock Advances in neural information processing systems. 2011;24.

\bibitem{mape_error_metric}
De~Myttenaere A, Golden B, Le~Grand B, Rossi F.
\newblock Mean absolute percentage error for regression models.
\newblock Neurocomputing. 2016;192:38-48.

\bibitem{GANs}
Creswell A, White T, Dumoulin V, Arulkumaran K, Sengupta B, Bharath AA.
\newblock Generative adversarial networks: An overview.
\newblock IEEE signal processing magazine. 2018;35(1):53-65.

\bibitem{ssm_pca_vs_autoencoder}
Ranjan A, Bolkart T, Sanyal S, Black MJ.
\newblock Generating 3D faces using convolutional mesh autoencoders.
\newblock In: Proceedings of the European Conference on Computer Vision (ECCV);
  2018. p. 704-20.

\bibitem{ML_autoencoders}
Wang Y, Yao H, Zhao S.
\newblock Auto-encoder based dimensionality reduction.
\newblock Neurocomputing. 2016;184:232-42.

\bibitem{ML_generation_of_synthetic}
Diller GP, Vahle J, Radke R, Vidal MLB, Fischer AJ, Bauer UM, et~al.
\newblock Utility of deep learning networks for the generation of artificial
  cardiac magnetic resonance images in congenital heart disease.
\newblock BMC Medical Imaging. 2020;20(1):1-8.

\bibitem{latin_hypercube_sampling}
Stein M.
\newblock Large sample properties of simulations using Latin hypercube
  sampling.
\newblock Technometrics. 1987;29(2):143-51.

\bibitem{CFD_ML_ships_pointnet}
Abbas A, Rafiee A, Haase M, Malcolm A.
\newblock Geometric Convolutional Neural Networks--A Journey to Surrogate
  Modelling of Maritime CFD.
\newblock In: The 9th Conference on Computational Methods in Marine Engineering
  (Marine 2021); 2022. .

\bibitem{pointnet}
Qi CR, Su H, Mo K, Guibas LJ.
\newblock PointNet: Deep Learning on Point Sets for 3D Classification and
  Segmentation.
\newblock CVPR 2017 Open Access. 2016.
\newblock Available from: \url{https://arxiv.org/abs/1612.00593}.

\bibitem{CFD_ML_autodesk_cars}
Umetani N, Bickel B.
\newblock {Learning Three-Dimensional Flow for Interactive Aerodynamic Design
  regression prediction for new shape}.
\newblock ACM Trans Graph. 2018;37(4).
\newblock Available from: \url{https://doi.org/10.1145/3197517.3201325}.

\bibitem{ML_Bayes_backprop}
Blundell C, Cornebise J, Kavukcuoglu K, Wierstra D.
\newblock Weight uncertainty in neural network.
\newblock In: International conference on machine learning. PMLR; 2015. p.
  1613-22.

\bibitem{cfd_geom_autoseg_javier_jcmr}
Montalt-Tordera J, Pajaziti E, Jones R, Sauvage E, Puranik R, Singh AAV,
  et~al.. Automatic Segmentation of the Great Arteries for Computational
  Hemodynamic Assessment. arXiv; 2022.
\newblock Available from: \url{https://arxiv.org/abs/2209.06117}.

\bibitem{cfd_geom_is_important_KwongTse}
Tse KM, Chang R, Lee HP, Lim SP, Venkatesh SK, Ho P.
\newblock A computational fluid dynamics study on geometrical influence of the
  aorta on haemodynamics.
\newblock European Journal of Cardio-Thoracic Surgery. 2013;43(4):829-38.

\bibitem{cfd_simple_models_useful}
{Accuracy vs. computational time: Translating aortic simulations to the
  clinic}.
\newblock Journal of Biomechanics. 2012;45(3):516-23.
\newblock Available from:
  \url{http://dx.doi.org/10.1016/j.jbiomech.2011.11.041}.

\bibitem{cfd_review_on_aorta_cfd}
Caballero AD, La{\'\i}n S.
\newblock A review on computational fluid dynamics modelling in human thoracic
  aorta.
\newblock Cardiovascular Engineering and Technology. 2013;4(2):103-30.

\bibitem{cfd_turbulence_coa}
Lantz J, Ebbers T, Engvall J, Karlsson M.
\newblock Numerical and experimental assessment of turbulent kinetic energy in
  an aortic coarctation.
\newblock Journal of biomechanics. 2013;46(11):1851-8.

\bibitem{cfd_inlet_idealised_isBad}
Youssefi P, Gomez A, Arthurs C, Sharma R, Jahangiri M, Alberto~Figueroa C.
\newblock Impact of patient-specific inflow velocity profile on hemodynamics of
  the thoracic aorta.
\newblock Journal of biomechanical engineering. 2018;140(1).

\bibitem{cfd_inlet_BC_Morbiducci_idealisedIsBad}
Morbiducci U, Ponzini R, Gallo D, Bignardi C, Rizzo G.
\newblock Inflow boundary conditions for image-based computational
  hemodynamics: impact of idealized versus measured velocity profiles in the
  human aorta.
\newblock Journal of biomechanics. 2013;46(1):102-9.

\bibitem{ssm_yevtushenko}
{Synthetic Database of Aortic Morphometry and Hemodynamics: Overcoming Medical
  Imaging Data Availability}.
\newblock IEEE Transactions on Medical Imaging. 2021;40(5):1438-49.

\bibitem{cfd_windkessel_overview}
Westerhof N, Lankhaar JW, Westerhof BE.
\newblock The arterial windkessel.
\newblock Medical \& biological engineering \& computing. 2009;47(2):131-41.

\bibitem{cfd_windkessel_study}
Romarowski RM, Lefieux A, Morganti S, Veneziani A, Auricchio F.
\newblock Patient-specific CFD modelling in the thoracic aorta with
  PC-MRI--based boundary conditions: A least-square three-element Windkessel
  approach.
\newblock International journal for numerical methods in biomedical
  engineering. 2018;34(11):e3134.

\bibitem{cfd_outlet_resistance_model_abdominal_aneurysm}
Les AS, Shadden SC, Figueroa CA, Park JM, Tedesco MM, Herfkens RJ, et~al.
\newblock Quantification of hemodynamics in abdominal aortic aneurysms during
  rest and exercise using magnetic resonance imaging and computational fluid
  dynamics.
\newblock Annals of biomedical engineering. 2010;38(4):1288-313.

\bibitem{cfd_inletoutlet_dont_affect_much}
Madhavan S, Kemmerling EMC.
\newblock The effect of inlet and outlet boundary conditions in image-based CFD
  modeling of aortic flow.
\newblock Biomedical engineering online. 2018;17(1):1-20.

\bibitem{cfd_aorta_BC_study}
Pirola S, Cheng Z, Jarral O, O'Regan D, Pepper J, Athanasiou T, et~al.
\newblock On the choice of outlet boundary conditions for patient-specific
  analysis of aortic flow using computational fluid dynamics.
\newblock Journal of biomechanics. 2017;60:15-21.

\end{thebibliography}

\end{document}